\documentclass[
 aip,
 apl,
 amsmath,amssymb,
 preprint,
 ]{revtex4-1}

\usepackage{graphicx}

\usepackage{color}

\usepackage[colorlinks=true,
        linkcolor=blue,
        citecolor=blue,
        filecolor=blue,
        urlcolor=blue,
        ]{hyperref}

\begin{document}

\title{Application of the equipartition theorem to the thermal excitation of 
         quartz tuning forks} 

\author{ Joachim Welker } 
\author{Frederico de Faria Elsner}
\author{Franz J. Giessibl}
\affiliation{Institute of Experimental and Applied Physics, Experimental
 Nanoscience, University of Regensburg, Universitaetsstrasse 31, 93053 
 Regensburg, Germany}

\date{\today}

\begin{abstract}
 The deflection signal of a thermally excited force sensor of an atomic force microscope can be analyzed 
 to gain important information about the detector noise and about the validity of the equipartion theorem of 
 thermodynamics. Here, we measured the
 temperature dependence of the thermal amplitude of a tuning fork and
 compared it to the expected values based on the equipartition theorem. In doing so,
 we prove the validity of these assumptions in the
 temperature range from 140$\,$K to 300$\,$K. Furthermore, the
 application of the equipartition theorem to quartz tuning forks 
 at liquid helium temperatures is discussed.
\end{abstract}


\maketitle 
Over the last decades quartz tuning forks have been used to build self-sensing
sensors in many research fields, for example hydrodynamics of quantum
fluids\cite{Clubb2004, Bradley2010}, spectroscopic gas
sensing\cite{Kosterev2005, Wojcik2006} and scanning probe
microscopy\cite{gunther1989, Karrai1995, Giessibl2000c}. 
In this paper we focus on quartz tuning forks used in frequency modulation atomic
force microscopy, although the results are also applicable to
other fields utilizing quartz tuning forks. Frequency modulation atomic
force microscopy (FM-AFM) with quartz tuning forks has put forth a
number of impressive results \cite{Giessibl2000a, Ternes2008, Gross2009}, e.g. FM-AFM was 
used to resolve the chemical structure of a molecule \cite{Gross2009}.
In FM-AFM the frequency shift $\Delta f$ of an oscillator measures the local
interaction of the microscope tip with the sample.
The force between tip and sample can be calculated from the frequency shift $\Delta
f$ if the sensor's resonance frequency, stiffness and oscillation
amplitude are known \cite{Giessibl2001, Sader2004}. 
Thus for determining relevant physical quantities out of the observed 
frequency shift those properties must be well-characterized.

A tuning fork is a cut piezoelectric quartz crystal with two prongs and gold electrodes along the prongs. 
When one or both prongs are deflected, charge accumulates on the electrodes. 
The sensitivity describes the relation between the piezoelectric output
signal and the deflection of a tuning fork. It is therefore essential to
know in order to determine the deflection amplitude. One method to
determine the sensitivity is to compare the output of the tuning fork due
to thermal excitation with the expected result based on the equipartition
theorem and the assumption that the first harmonic mode is the only
mode significantly excited.
The equipartition theorem is also used in FM-AFM 
to calculate the fundamental noise limits in force detection  due
to thermal excitation\cite{Albrecht1991, Colchero2011}. 
Understanding the fundamental noise limits is very
important for judging and improving a system's performance. 
In this paper we show the validity of using the equipartition theorem in
the temperature range from 140 K to 300 K and discuss its application at
liquid helium temperatures.

The equipartition theorem states that each degree of freedom holds a
thermal energy of $\frac{1}{2}\,k_BT$, where $k_B$ is Boltzmann's
constant and $T$ is the temperature in Kelvin. For a coupled oscillator
like the tuning fork with one degree of freedom this leads to the
relation 
\begin{equation}
  2 \cdot \frac{1}{2} k  \left(A_{th}^{EqT}\right)^2=\frac{1}{2}k_BT  
    \quad \Rightarrow \quad
    A_{th}^{EqT}=\sqrt{\frac{k_{B}T }{2k}} \quad,
    \label{eq:A_th_EqT}
\end{equation}
where $k$ is the spring constant and $A_{th}^{EqT}$ is the  thermal deflection
amplitude of one prong. 

Experimentally,
the piezoelectric signal of a tuning fork is measured with a
transimpedance amplifier. The voltage output $V_{th}$ of the
transimpedance amplifier can be converted to the thermal deflection
amplitude $A_{th}^{Exp}$ by knowing the sensitivity $S$ of the quartz tuning
fork: 
\begin{equation}
   A_{th}^{Exp}=S^{-1}\cdot V_{th} \quad.
   \label{eq:A_th_Exp}
\end{equation}
In this notation the sensitivity depends on the amplifier.
The theoretical sensitivity $S$ of a tuning fork calculated with beam
theory\cite{Giessibl2000c} is given by 
\begin{equation}
   S= 2\cdot g\cdot 2.8\,\mu\text{C} / \text{m}\cdot 2 \pi f_0 \cdot R \cdot
      G_{f_0} \quad,
   \label{eq:S}
\end{equation}
where $f_0$ is the resonance frequency and $R$ is the feedback resistor
of the transimpedance amplifier. Compared with the equation given in
reference \onlinecite{Giessibl2000c}, the two dimensionless factors $G_{f_0}$ and $g$
have been added to account for the limited bandwidth of the transimpedance 
amplifier and the geometrical configuration of the tuning fork's
electrodes. The geometry factor is independent of the amplifier and reduces 
the generated charge per deflection by the factor $g=0.51$. It was determined by a strobe 
microscopy deflection measurement similar to
that reported in reference \onlinecite{Bradley2010}.

In our experiments we use an encapsulated tuning fork with a resonance
frequency of $f_0=32768\,$Hz,
a quality factor of $Q=30\times10^3$,
and a spring constant of $k=1800\,\rm{N}/\rm{m}$.
This tuning fork is mounted into a bore in a metal slab that serves as
a thermal mass. The metal slab with the tuning fork is first immersed in
liquid nitrogen until it is thermalized, which is indicated by the end of
the heavy boiling of the liquid nitrogen. Subsequently, the cold metal
slab is put inside polystyrene insulation. As the metal slab warms up,
the tuning fork's output is continually being measured. The generated
current is converted to a voltage by a transimpedance amplifier
consisting of an op-amp (AD823) with a 100$\,\rm{M}\Omega$ feedback
resistor.
The op-amp and feedback resistor are both kept at room temperature. The 
transimpedance gain of $100\,\rm{mV}/\rm{nA}$ of the amplifier is
reduced by the factor $G_{f_0}=0.063$ due to its limited bandwidth of $3\,\rm{kHz}$.
The output of the transimpedance amplifier is fed into a
spectrum analyzer (SRS SR760) where the power spectral density $n^2_{V}(f)$
is recorded. 

\begin{figure}
  \includegraphics{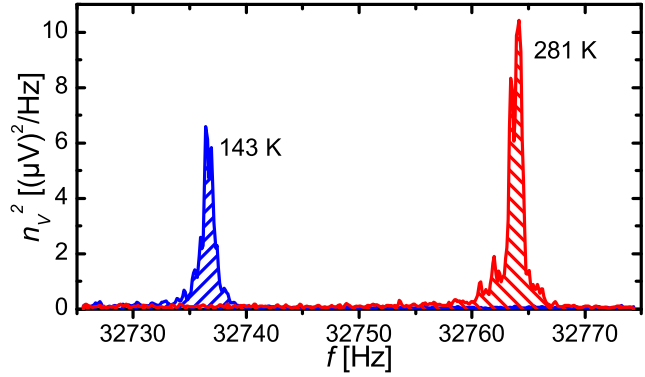}
  \caption{\label{pic:psd_vs_freq}(Color online) Power spectral density 
            $n^2_{V}$ of a tuning fork at $143\,$K (blue) and 281$\,$K (red)
             due to the intrinsic thermal excitation. The area 
            under the curves corresponds to the squared voltage output 
            generated by the thermal deflection amplitude.}
\end{figure}
In figure \ref{pic:psd_vs_freq} two spectra of the power spectral density  
are shown at two different
temperatures. The temperature $T$ in Kelvin of each spectrum is
determined by the shift of the actual resonance frequency $f$ with respect
to the resonance frequency $f_0$\cite{Microcrystal2010}: 
$T=298.15\text{K}\pm 5345\,\text{K}\sqrt{-\Delta f / f_0}$.
It should be noted that the frequency shift  $\Delta f = f-f_0$ is always 
negative off the resonance. This dependence was previously verified for a 
tuning fork in the temperature range from 150$\,$K to 300$\,$K \cite{Hembacher2002}.
The power spectral density $n^2_{V}(f)$ has two components $n_{th}^2$ and
$n^2_{el}$. 
The density $n_{th}$ is the contribution of the thermal energy to the
deflection of the tuning fork, whereas $n_{el}$ is the electrical noise
density of the transimpedance amplifier. 
The relevant output is the area under the resonance peak $n_{th}^{2}$
without the contribution of the electrical noise density $n^2_{el}$. Thus
the squared voltage output of the thermally excited tuning fork can be
expressed as 
\begin{equation}
 V_{th}^2=  \int_{f_0-B/2}^{f_0+B/2}{n_{th}^2\left(f\right) \text{d}f}=
          \int_{f_0-B/2}^{f_0+B/2}{\left( n^2_{V}\left(f\right) -n_{el}^{2}\right) 
           \text{d}f}\ .
 \label{eq:V_th}
\end{equation}
Here the electrical noise density $n_{el}$ is assumed to be white over
the bandwidth $B$, which is the local frequency range around the thermal
peak. Experimentally, a bandwidth of $48.75\,\rm{Hz}$ and a center 
frequency of $32750\,\rm{Hz}$ are used for all measurements as shown in
figure \ref{pic:psd_vs_freq}.
The electrical noise density $n_{el}$ corresponds to the baseline of the
density $n_{V}$. It can be estimated by averaging $n_{V}$ away from the
thermal peak which leads to a value of $n_{el}\,\approx
300\,\rm{nV}/\sqrt{\rm{Hz}}$ \cite{Note1}.
 
\begin{figure}
  \includegraphics{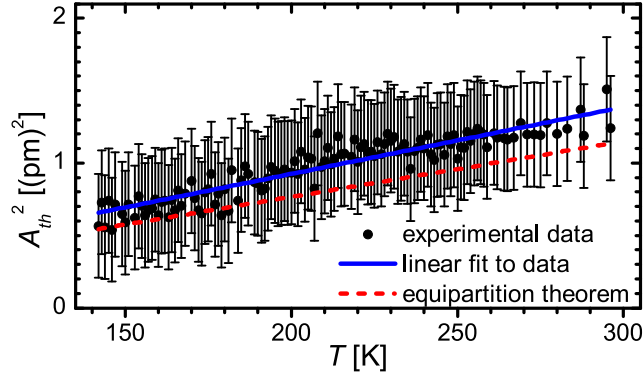}
  \caption{\label{pic:A_th_vs_temp}(Color online) Temperature dependence
   of the tuning
   fork's thermal amplitude $A_{th}$. The solid, blue line shows the
   linear behavior in $A_{th}^2$ as a result of the equipartition
   theorem. The dashed, red line shows the theoretical dependence
   according to equation \eqref{eq:A_th_EqT} with no adjustable
   parameter. }
\end{figure}
Figure \ref{pic:A_th_vs_temp} shows the temperature dependence of the
squared thermal deflection amplitude. The thermal deflection amplitudes
are determined out of the temperature dependent spectral noise densities
$n^2_{V}(f)$ using equations \eqref{eq:A_th_Exp}, \eqref{eq:S} and
\eqref{eq:V_th}. The measurements are averaged in 1$\,$K intervals and
the error bars show the maximum standard deviation. The dashed, red line
depicts the expected thermal deflection amplitude according to the 
equipartition theorem, equation \eqref{eq:A_th_EqT}. It has to  be noted 
that there are no free parameters neither in the measurements of $A_{th}^{Exp}$
nor in the calculation of $A_{th}^{EqT}$ used in figure 
\ref{pic:A_th_vs_temp}. The experimental data resembles the expected 
linear temperature dependence in $A_{th}^2$, as shown by the solid, 
blue linear fit without offset. 
However, there is a slight deviation in 
the slopes between the linear fit and the theoretical temperature 
dependence. We propose that this can be explained by an inaccuracy 
in the sensitivity $S$. The sensitivity according to equation \eqref{eq:S}
results in $S=3.74\,\rm{\mu V}/\rm{pm}$ at the resonance
frequency. If the sensitivity is assumed to be a free parameter, the
optimal sensitivity to fit the data to the equipartition theorem is
$S^{EqT}=4.09\,\rm{\mu V}/\rm{pm}$, which is a 9\% change. The change in
the sensitivity due to frequency shift can be ignored as the frequency
changes with temperature less than 0.1\%.
However, the qualitative and quantitative agreement proves within the
measurement accuracy the validity of using the equipartition theorem for
thermally excited tuning forks.

Of course, the above argument only holds, if the tuning fork is solely
excited by thermal energy. It must be ensured that there are no
mechanical excitations due to vibrations. In order to compare mechanical
noise with the thermal noise the equivalent white noise drive\cite{Colchero2011} 
\begin{equation}
  \alpha_{th}=\sqrt{ 2k_BT /(\pi f_0 k Q)}
\end{equation}
can be used. The white noise drive given in $\rm{m}/\sqrt{\rm{Hz}}$ describes 
an equivalent mechanical drive of a harmonic oscillator in accordance with 
the equipartition theorem. 

In general, tuning forks are very robust against vibrations of the base plate
because vibrations would lead to a symmetric oscillation mode of the prongs, 
whereas the tuning fork oscillates preferably in the anti-symmetric mode. 
In the anti-symmetric oscillation mode the tuning fork's center of mass  
stays at rest leading to less dissipation in the baseplate. Furthermore, charges 
produced by the symmetric mode cancel out each other due to the asymmetric 
electrode configuration, which suppresses the symmetric mode.

In FM-AFM tuning forks are often used in the qPlus
configuration\cite{Giessibl2000c}, where one prong of the tuning fork is
glued to a massive substrate. Though the qPlus sensor has proven itself \cite{Giessibl2000a, Ternes2008, Gross2009},
it is more sensitive to vibrations of the baseplate. This plays a serious
role especially at low temperatures. At liquid helium temperatures the
quality factor $Q$ of the qPlus sensor can raise up to $2\cdot10^5$.
This means that even very small vibrations of the baseplate can notably
excite the qPlus sensor. For example, the theoretical thermal amplitude
at 4.4$\,$K is $A_{th}^{EqT}=184\,$fm 
according to a thermal white noise drive  of 
$\alpha_{th}=2.56 \, \rm{am} / \sqrt{\rm{Hz}}$. 
With a quality factor of
$Q=1\cdot10^5$ any vibration bigger than 1.84$\,$am (!) would result in
a deflection amplitude bigger than the thermal deflection amplitude.
If the tuning fork is mounted on a dither piezo for mechanical excitation, 
the vibrations can result from noise in the excitation voltage applied to 
the piezo. The white noise drive $\alpha_{th}$ can be compared with this 
electrical noise knowing the sensitivity of the dither piezo. In our 
Omicron LT qPlus AFM/STM
the  white  noise drive  
$\alpha_{th}=2.56 \, \rm{am} / \sqrt{\rm{Hz}}$ corresponds to
an electrical noise density $n_{el}^{\rm{\it{Piezo}}}=1.94\, \rm{nV}/\sqrt{\rm{Hz}}$
of the excitation voltage. 
Thus in this setup the noise on the excitation voltage needs to be better than
$0.2\, \rm{nV}/\sqrt{\rm{Hz}}$  in order to justifiably exploit the equipartition 
theorem for determining the thermal deflection amplitude at liquid helium temperatures.

\begin{figure}
	\includegraphics{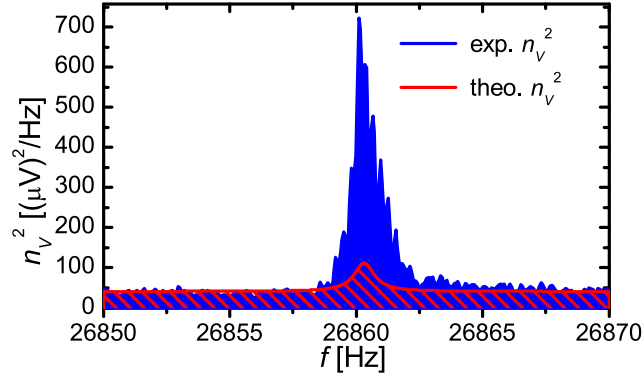}
	\caption{(Color online) Experimental and theoretical power spectral densities 
	          of a qPlus sensor at a microscope temperature of 4.4$\,$K. The 
	          theoretical power spectral density is modeled by Lorentzian function 
	          with the experimental noise floor. Due to mechanical noise in the system
	          the experimental power spectral density results in an effective 
	          temperature of 32.5$\,$K.}
	\label{fig:eff_thermal_spec}
\end{figure}

Conversely,
if this stability is not provided, the equipartition theorem can be used
to assign an effective temperature $T_{\text{\it{eff}}}$ to the qPlus
sensor: 
\begin{equation}
	\centering
   T_{\text{\it{eff}}}=\frac{k}{k_B}\left(A^{Exp}_{noise}\right)^2  \quad.
   \label{eq:T_eff}
\end{equation}
The amplitude $A^{Exp}_{noise}$ is determined like the thermal amplitude
$A^{Exp}_{th}$ before by recording the power spectral spectral density
$n^2_{V}(f)$ without active driving and using equations 
\eqref{eq:A_th_Exp} and \eqref{eq:V_th}. 
This effective temperature can be significantly higher than the temperature 
of the thermal bath. Figure \ref{fig:eff_thermal_spec} shows a spectrum of 
qPlus sensor measured in our Omicron LT qPlus AFM/STM at 4.4$\,$K without 
active driving and the piezos grounded. The sensitivity of the sensor was 
determined in situ with a tunneling current controlled amplitude 
determination\cite{Simon2007} resulting in $S=56.7\,\mu\rm{V}/\rm{pm}$.
The deflection amplitude is given by $A^{Exp}_{noise}=499\,\rm{fm}$, 
which corresponds to an effective temperature of $T_{\text{\it{eff}}}=32.5\,$K.
Therefore also the frequency noise $\delta f_{ T_{\text{\it{eff}}}}$
and the minimum detectable force gradient $\delta k_{ T_{\text{\it{eff}}}}$ 
rise according to equations (18) and (19) in reference \onlinecite{Albrecht1991}.

In summary, we have shown the temperature dependence of the thermal
deflection amplitude of a quartz tuning fork. This dependence shows
clearly a linear behavior in $A_{th}^2$ as expected by the equipartition theorem.
Furthermore, the application of the equipartition theorem to qPlus sensors
at liquid helium temperatures was discussed. It was shown that the mechanical
stability has to be in the order of attometers in order to gain 
thermal-noise-limited performance. But the equipartition theorem can 
also be used to determine a realistic minimum detectable force gradient 
for a given setup.
 
We acknowledge Florian Pielmeier, Thomas Hofmann and Alfred J.
Weymouth for fruitful discussion and support.


\begin{thebibliography}{18}%
\makeatletter
\providecommand \@ifxundefined [1]{%
 \@ifx{#1\undefined}
}%
\providecommand \@ifnum [1]{%
 \ifnum #1\expandafter \@firstoftwo
 \else \expandafter \@secondoftwo
 \fi
}%
\providecommand \@ifx [1]{%
 \ifx #1\expandafter \@firstoftwo
 \else \expandafter \@secondoftwo
 \fi
}%
\providecommand \natexlab [1]{#1}%
\providecommand \enquote  [1]{``#1''}%
\providecommand \bibnamefont  [1]{#1}%
\providecommand \bibfnamefont [1]{#1}%
\providecommand \citenamefont [1]{#1}%
\providecommand \href@noop [0]{\@secondoftwo}%
\providecommand \href [0]{\begingroup \@sanitize@url \@href}%
\providecommand \@href[1]{\@@startlink{#1}\@@href}%
\providecommand \@@href[1]{\endgroup#1\@@endlink}%
\providecommand \@sanitize@url [0]{\catcode `\\12\catcode `\$12\catcode
  `\&12\catcode `\#12\catcode `\^12\catcode `\_12\catcode `\%12\relax}%
\providecommand \@@startlink[1]{}%
\providecommand \@@endlink[0]{}%
\providecommand \url  [0]{\begingroup\@sanitize@url \@url }%
\providecommand \@url [1]{\endgroup\@href {#1}{\urlprefix }}%
\providecommand \urlprefix  [0]{URL }%
\providecommand \Eprint [0]{\href }%
\providecommand \doibase [0]{http://dx.doi.org/}%
\providecommand \selectlanguage [0]{\@gobble}%
\providecommand \bibinfo  [0]{\@secondoftwo}%
\providecommand \bibfield  [0]{\@secondoftwo}%
\providecommand \translation [1]{[#1]}%
\providecommand \BibitemOpen [0]{}%
\providecommand \bibitemStop [0]{}%
\providecommand \bibitemNoStop [0]{.\EOS\space}%
\providecommand \EOS [0]{\spacefactor3000\relax}%
\providecommand \BibitemShut  [1]{\csname bibitem#1\endcsname}%
\let\auto@bib@innerbib\@empty
%</preamble>
\bibitem [{\citenamefont {Clubb}\ \emph {et~al.}(2004)\citenamefont {Clubb},
  \citenamefont {Buu}, \citenamefont {Bowley}, \citenamefont {Nyman},\ and\
  \citenamefont {Owers-Bradley}}]{Clubb2004}%
  \BibitemOpen
  \bibfield  {author} {\bibinfo {author} {\bibfnamefont {D.~O.}\ \bibnamefont
  {Clubb}}, \bibinfo {author} {\bibfnamefont {O.~V.~L.}\ \bibnamefont {Buu}},
  \bibinfo {author} {\bibfnamefont {R.~M.}\ \bibnamefont {Bowley}}, \bibinfo
  {author} {\bibfnamefont {R.}~\bibnamefont {Nyman}}, \ and\ \bibinfo {author}
  {\bibfnamefont {J.~R.}\ \bibnamefont {Owers-Bradley}},\ }\href {\doibase
  10.1023/B:JOLT.0000035368.63197.16} {\bibfield  {journal} {\bibinfo
  {journal} {J. Low Temp. Phys.}\ }\textbf {\bibinfo {volume} {136}},\ \bibinfo
  {pages} {1} (\bibinfo {year} {2004})}\BibitemShut {NoStop}%
\bibitem [{\citenamefont {Bradley}\ \emph {et~al.}(2010)\citenamefont
  {Bradley}, \citenamefont {Crookston}, \citenamefont {Fear}, \citenamefont
  {Fisher}, \citenamefont {Foulds}, \citenamefont {Garg}, \citenamefont
  {Gu\'{e}nault}, \citenamefont {Guise}, \citenamefont {Haley}, \citenamefont
  {Kolosov}, \citenamefont {Pickett}, \citenamefont {Schanen},\ and\
  \citenamefont {Tsepelin}}]{Bradley2010}%
  \BibitemOpen
  \bibfield  {author} {\bibinfo {author} {\bibfnamefont {D.~I.}\ \bibnamefont
  {Bradley}}, \bibinfo {author} {\bibfnamefont {P.}~\bibnamefont {Crookston}},
  \bibinfo {author} {\bibfnamefont {M.~J.}\ \bibnamefont {Fear}}, \bibinfo
  {author} {\bibfnamefont {S.~N.}\ \bibnamefont {Fisher}}, \bibinfo {author}
  {\bibfnamefont {G.}~\bibnamefont {Foulds}}, \bibinfo {author} {\bibfnamefont
  {D.}~\bibnamefont {Garg}}, \bibinfo {author} {\bibfnamefont {A.~M.}\
  \bibnamefont {Gu\'{e}nault}}, \bibinfo {author} {\bibfnamefont
  {E.}~\bibnamefont {Guise}}, \bibinfo {author} {\bibfnamefont {R.~P.}\
  \bibnamefont {Haley}}, \bibinfo {author} {\bibfnamefont {O.}~\bibnamefont
  {Kolosov}}, \bibinfo {author} {\bibfnamefont {G.~R.}\ \bibnamefont
  {Pickett}}, \bibinfo {author} {\bibfnamefont {R.}~\bibnamefont {Schanen}}, \
  and\ \bibinfo {author} {\bibfnamefont {V.}~\bibnamefont {Tsepelin}},\ }\href
  {\doibase 10.1007/s10909-010-0227-y} {\bibfield  {journal} {\bibinfo
  {journal} {J. Low Temp. Phys.}\ }\textbf {\bibinfo {volume} {161}},\ \bibinfo
  {pages} {536} (\bibinfo {year} {2010})}\BibitemShut {NoStop}%
\bibitem [{\citenamefont {Kosterev}\ \emph {et~al.}(2005)\citenamefont
  {Kosterev}, \citenamefont {Tittel}, \citenamefont {Serebryakov},
  \citenamefont {Malinovsky},\ and\ \citenamefont {Morozov}}]{Kosterev2005}%
  \BibitemOpen
  \bibfield  {author} {\bibinfo {author} {\bibfnamefont {A.~A.}\ \bibnamefont
  {Kosterev}}, \bibinfo {author} {\bibfnamefont {F.~K.}\ \bibnamefont
  {Tittel}}, \bibinfo {author} {\bibfnamefont {D.~V.}\ \bibnamefont
  {Serebryakov}}, \bibinfo {author} {\bibfnamefont {A.~L.}\ \bibnamefont
  {Malinovsky}}, \ and\ \bibinfo {author} {\bibfnamefont {I.~V.}\ \bibnamefont
  {Morozov}},\ }\href {\doibase 10.1063/1.1884196} {\bibfield  {journal}
  {\bibinfo  {journal} {Rev. Sci. Inst.}\ }\textbf {\bibinfo {volume} {76}},\
  \bibinfo {pages} {043105} (\bibinfo {year} {2005})}\BibitemShut {NoStop}%
\bibitem [{\citenamefont {Wojcik}\ \emph {et~al.}(2006)\citenamefont {Wojcik},
  \citenamefont {Phillips}, \citenamefont {Cannon},\ and\ \citenamefont
  {Taubman}}]{Wojcik2006}%
  \BibitemOpen
  \bibfield  {author} {\bibinfo {author} {\bibfnamefont {M.~D.}\ \bibnamefont
  {Wojcik}}, \bibinfo {author} {\bibfnamefont {M.~C.}\ \bibnamefont
  {Phillips}}, \bibinfo {author} {\bibfnamefont {B.~D.}\ \bibnamefont
  {Cannon}}, \ and\ \bibinfo {author} {\bibfnamefont {M.~S.}\ \bibnamefont
  {Taubman}},\ }\href {\doibase 10.1007/s00340-006-2394-8} {\bibfield
  {journal} {\bibinfo  {journal} {Appl. Phys. B}\ }\textbf {\bibinfo {volume}
  {85}},\ \bibinfo {pages} {307} (\bibinfo {year} {2006})}\BibitemShut
  {NoStop}%
\bibitem [{\citenamefont {G\"{u}nther}, \citenamefont {Fischer},\ and\
  \citenamefont {Dransfeld}(1989)}]{gunther1989}%
  \BibitemOpen
  \bibfield  {author} {\bibinfo {author} {\bibfnamefont {P.}~\bibnamefont
  {G\"{u}nther}}, \bibinfo {author} {\bibfnamefont {U.~C.}\ \bibnamefont
  {Fischer}}, \ and\ \bibinfo {author} {\bibfnamefont {K.}~\bibnamefont
  {Dransfeld}},\ }\href {\doibase 10.1007/BF00694423} {\bibfield  {journal}
  {\bibinfo  {journal} {Appl. Phys. B}\ }\textbf {\bibinfo {volume} {48}},\
  \bibinfo {pages} {89} (\bibinfo {year} {1989})}\BibitemShut {NoStop}%
\bibitem [{\citenamefont {Karrai}\ and\ \citenamefont
  {Grober}(1995)}]{Karrai1995}%
  \BibitemOpen
  \bibfield  {author} {\bibinfo {author} {\bibfnamefont {K.}~\bibnamefont
  {Karrai}}\ and\ \bibinfo {author} {\bibfnamefont {R.~D.}\ \bibnamefont
  {Grober}},\ }\href {\doibase 10.1063/1.113340} {\bibfield  {journal}
  {\bibinfo  {journal} {Appl. Phys. Lett.}\ }\textbf {\bibinfo {volume} {66}},\
  \bibinfo {pages} {1842} (\bibinfo {year} {1995})}\BibitemShut {NoStop}%
\bibitem [{\citenamefont {Giessibl}(2000)}]{Giessibl2000c}%
  \BibitemOpen
  \bibfield  {author} {\bibinfo {author} {\bibfnamefont {F.~J.}\ \bibnamefont
  {Giessibl}},\ }\href {\doibase 10.1063/1.126067} {\bibfield  {journal}
  {\bibinfo  {journal} {Appl. Phys. Lett.}\ }\textbf {\bibinfo {volume} {76}},\
  \bibinfo {pages} {1470} (\bibinfo {year} {2000})}\BibitemShut {NoStop}%
\bibitem [{\citenamefont {Giessibl}\ \emph {et~al.}(2000)\citenamefont
  {Giessibl}, \citenamefont {Hembacher}, \citenamefont {Bielefeldt},\ and\
  \citenamefont {Mannhart}}]{Giessibl2000a}%
  \BibitemOpen
  \bibfield  {author} {\bibinfo {author} {\bibfnamefont {F.~J.}\ \bibnamefont
  {Giessibl}}, \bibinfo {author} {\bibfnamefont {S.}~\bibnamefont {Hembacher}},
  \bibinfo {author} {\bibfnamefont {H.}~\bibnamefont {Bielefeldt}}, \ and\
  \bibinfo {author} {\bibfnamefont {J.}~\bibnamefont {Mannhart}},\ }\href
  {\doibase 10.1126/science.289.5478.422} {\bibfield  {journal} {\bibinfo
  {journal} {Science}\ }\textbf {\bibinfo {volume} {289}},\ \bibinfo {pages}
  {422} (\bibinfo {year} {2000})}\BibitemShut {NoStop}%
\bibitem [{\citenamefont {Ternes}\ \emph {et~al.}(2008)\citenamefont {Ternes},
  \citenamefont {Lutz}, \citenamefont {Hirjibehedin}, \citenamefont
  {Giessibl},\ and\ \citenamefont {Heinrich}}]{Ternes2008}%
  \BibitemOpen
  \bibfield  {author} {\bibinfo {author} {\bibfnamefont {M.}~\bibnamefont
  {Ternes}}, \bibinfo {author} {\bibfnamefont {C.~P.}\ \bibnamefont {Lutz}},
  \bibinfo {author} {\bibfnamefont {C.~F.}\ \bibnamefont {Hirjibehedin}},
  \bibinfo {author} {\bibfnamefont {F.~J.}\ \bibnamefont {Giessibl}}, \ and\
  \bibinfo {author} {\bibfnamefont {A.~J.}\ \bibnamefont {Heinrich}},\ }\href
  {\doibase 10.1126/science.1150288} {\bibfield  {journal} {\bibinfo  {journal}
  {Science}\ }\textbf {\bibinfo {volume} {319}},\ \bibinfo {pages} {1066}
  (\bibinfo {year} {2008})}\BibitemShut {NoStop}%
\bibitem [{\citenamefont {Gross}\ \emph {et~al.}(2009)\citenamefont {Gross},
  \citenamefont {Mohn}, \citenamefont {Moll}, \citenamefont {Liljeroth},\ and\
  \citenamefont {Meyer}}]{Gross2009}%
  \BibitemOpen
  \bibfield  {author} {\bibinfo {author} {\bibfnamefont {L.}~\bibnamefont
  {Gross}}, \bibinfo {author} {\bibfnamefont {F.}~\bibnamefont {Mohn}},
  \bibinfo {author} {\bibfnamefont {N.}~\bibnamefont {Moll}}, \bibinfo {author}
  {\bibfnamefont {P.}~\bibnamefont {Liljeroth}}, \ and\ \bibinfo {author}
  {\bibfnamefont {G.}~\bibnamefont {Meyer}},\ }\href {\doibase
  10.1126/science.1176210} {\bibfield  {journal} {\bibinfo  {journal}
  {Science}\ }\textbf {\bibinfo {volume} {325}},\ \bibinfo {pages} {1110}
  (\bibinfo {year} {2009})}\BibitemShut {NoStop}%
\bibitem [{\citenamefont {Giessibl}(2001)}]{Giessibl2001}%
  \BibitemOpen
  \bibfield  {author} {\bibinfo {author} {\bibfnamefont {F.~J.}\ \bibnamefont
  {Giessibl}},\ }\href {\doibase 10.1063/1.1335546} {\bibfield  {journal}
  {\bibinfo  {journal} {Appl. Phys. Lett.}\ }\textbf {\bibinfo {volume} {78}},\
  \bibinfo {pages} {123} (\bibinfo {year} {2001})}\BibitemShut {NoStop}%
\bibitem [{\citenamefont {Sader}\ and\ \citenamefont
  {Jarvis}(2004)}]{Sader2004}%
  \BibitemOpen
  \bibfield  {author} {\bibinfo {author} {\bibfnamefont {J.~E.}\ \bibnamefont
  {Sader}}\ and\ \bibinfo {author} {\bibfnamefont {S.~P.}\ \bibnamefont
  {Jarvis}},\ }\href {\doibase 10.1063/1.1667267} {\bibfield  {journal}
  {\bibinfo  {journal} {Appl. Phys. Lett.}\ }\textbf {\bibinfo {volume} {84}},\
  \bibinfo {pages} {1801} (\bibinfo {year} {2004})}\BibitemShut {NoStop}%
\bibitem [{\citenamefont {Albrecht}\ \emph {et~al.}(1991)\citenamefont
  {Albrecht}, \citenamefont {Grutter}, \citenamefont {Horne},\ and\
  \citenamefont {Rugar}}]{Albrecht1991}%
  \BibitemOpen
  \bibfield  {author} {\bibinfo {author} {\bibfnamefont {T.~R.}\ \bibnamefont
  {Albrecht}}, \bibinfo {author} {\bibfnamefont {P.}~\bibnamefont {Grutter}},
  \bibinfo {author} {\bibfnamefont {D.}~\bibnamefont {Horne}}, \ and\ \bibinfo
  {author} {\bibfnamefont {D.}~\bibnamefont {Rugar}},\ }\href
  {http://dx.doi.org/10.1063/1.347347} {\bibfield  {journal} {\bibinfo
  {journal} {J. Appl. Phys.}\ }\textbf {\bibinfo {volume} {69}},\ \bibinfo
  {pages} {668} (\bibinfo {year} {1991})}\BibitemShut {NoStop}%
\bibitem [{\citenamefont {Colchero}\ \emph {et~al.}(2011)\citenamefont
  {Colchero}, \citenamefont {Cuenca}, \citenamefont {Mart\'inez}, \citenamefont
  {Abad}, \citenamefont {Garc\'ia}, \citenamefont {Palacios-Lid\'on},\ and\
  \citenamefont {Abell\'an}}]{Colchero2011}%
  \BibitemOpen
  \bibfield  {author} {\bibinfo {author} {\bibfnamefont {J.}~\bibnamefont
  {Colchero}}, \bibinfo {author} {\bibfnamefont {M.}~\bibnamefont {Cuenca}},
  \bibinfo {author} {\bibfnamefont {J.~F.~G.}\ \bibnamefont {Mart\'inez}},
  \bibinfo {author} {\bibfnamefont {J.}~\bibnamefont {Abad}}, \bibinfo {author}
  {\bibfnamefont {B.~P.}\ \bibnamefont {Garc\'ia}}, \bibinfo {author}
  {\bibfnamefont {E.}~\bibnamefont {Palacios-Lid\'on}}, \ and\ \bibinfo
  {author} {\bibfnamefont {J.}~\bibnamefont {Abell\'an}},\ }\href {\doibase
  10.1063/1.3533769} {\bibfield  {journal} {\bibinfo  {journal} {J. Appl.
  Phys.}\ }\textbf {\bibinfo {volume} {109}},\ \bibinfo {pages} {024310}
  (\bibinfo {year} {2011})}\BibitemShut {NoStop}%
\bibitem [{Mic(2010)}]{Microcrystal2010}%
  \BibitemOpen
  \href
  {http://www.microcrystal.com/CMSPages/GetFile.aspx?nodeguid=809bc04c-68c6-4aa9-9ebb-7f88474f8f26}
  {\emph {\bibinfo {title} {DS-Series Watch Crystal 32.768 kHz}}},\ \bibinfo
  {organization} {Micro Crystal AG},\ \bibinfo {address} {M\"{u}hlestrasse 14,
  CH-2540 Grenchen, Switzerland} (\bibinfo {year} {2010})\BibitemShut {NoStop}%
\bibitem [{\citenamefont {Hembacher}, \citenamefont {Giessibl},\ and\
  \citenamefont {Mannhart}(2002)}]{Hembacher2002}%
  \BibitemOpen
  \bibfield  {author} {\bibinfo {author} {\bibfnamefont {S.}~\bibnamefont
  {Hembacher}}, \bibinfo {author} {\bibfnamefont {F.~J.}\ \bibnamefont
  {Giessibl}}, \ and\ \bibinfo {author} {\bibfnamefont {J.}~\bibnamefont
  {Mannhart}},\ }\href {\doibase 10.1016/S0169-4332(01)00976-X} {\bibfield
  {journal} {\bibinfo  {journal} {Appl. Surf. Sci.}\ }\textbf {\bibinfo
  {volume} {188}},\ \bibinfo {pages} {445} (\bibinfo {year}
  {2002})}\BibitemShut {NoStop}%
\bibitem [{Note1()}]{Note1}%
  \BibitemOpen
  \bibinfo {note} {The electrical noise density is smaller than the Johnson
  noise of the feedback resistor ($1.3\protect \tmspace +\thinmuskip
  {.1667em}\mu \protect \rm {V}/\protect \sqrt {\protect \rm {Hz}}$), because
  the Johnson noise is attenuated due to the limited bandwidth of the
  amplifier.}\BibitemShut {Stop}%
\bibitem [{\citenamefont {Simon}, \citenamefont {Heyde},\ and\ \citenamefont
  {Rust}(2007)}]{Simon2007}%
  \BibitemOpen
  \bibfield  {author} {\bibinfo {author} {\bibfnamefont {G.~H.}\ \bibnamefont
  {Simon}}, \bibinfo {author} {\bibfnamefont {M.}~\bibnamefont {Heyde}}, \ and\
  \bibinfo {author} {\bibfnamefont {H.}~\bibnamefont {Rust}},\ }\href {\doibase
  10.1088/0957-4484/18/25/255503} {\bibfield  {journal} {\bibinfo  {journal}
  {Nanotechnology}\ }\textbf {\bibinfo {volume} {18}},\ \bibinfo {pages}
  {255503} (\bibinfo {year} {2007})}\BibitemShut {NoStop}%
\end{thebibliography}
\end{document}